\documentclass{ar-1col-S2O}

\usepackage{amssymb}

\usepackage[comma]{natbib}
\usepackage{url}
\usepackage{setspace}
\usepackage{amsmath}

\newcommand{\sized}{D}

\jname{Xxxx. Xxx. Xxx. Xxx.}
\jvol{AA}
\jyear{2023}
\doi{10.1146/((please add article doi))}

\begin{document}

\markboth{Rui Ni}{Deformation and Breakup in Turbulence}

\title{Deformation and breakup of bubbles and drops in turbulence}

\author{Rui Ni$^1$
\affil{$^1$Department of Mechanical Engineering, The Johns Hopkins University, Baltimore, MD 21218, USA; email: rui.ni@jhu.edu}}

\begin{abstract}


Fragmentation of bubbles and droplets in turbulence produces a dispersed phase spanning a broad range of scales, encompassing everything from droplets in nanoemulsions to centimeter-sized bubbles entrained in breaking waves. Along with deformation, fragmentation plays a crucial role in enhancing interfacial area, with far-reaching implications across various industries, including food, pharmaceuticals, and ocean engineering. However, understanding and modeling these processes is challenging due to the complexity of anisotropic and inhomogeneous turbulence typically involved, the unknown residence time in regions with different turbulence intensities, and difficulties arising from the density and viscosity ratios. Despite these challenges, recent advances have provided new insights into the underlying physics of deformation and fragmentation  in turbulence. This review summarizes existing works in various fields, highlighting key results and uncertainties, and examining the impact on turbulence modulation, drag reduction, and heat and mass transfer. 
\end{abstract}

\maketitle




\section{Introduction}

Mixing two immiscible fluids (gas-liquid or liquid-liquid) in turbulence produces polydispersed droplets or bubbles that can freely deform, break, and coalesce while interacting with the surrounding turbulence. These processes are fundamentally important and practically relevant to multiple fields, including bubble-mediated air-sea mass exchange \citep{villermaux2022bubbles}, chemical emulsions, food science, nuclear thermal hydraulics, and two-phase heat transfer. In contrast to the deformation and breakup of droplets in low-Reynolds-number viscous flows \citep{stone1994dynamics}, in turbulence, these dynamics are intimately linked to multiple length and time scales associated with the background turbulent eddies. 

A wide range of drop/bubble sizes can therefore be achieved via the adjustment of the turbulence characteristics. For example, turbulence generated by a simple batch stirrer system can break an oil-water mixture into macroemulsions with the size of the dispersed oil droplets at $\mathcal{O}(1$--$100$ $\mu$m). But if nanoemulsions with droplets of $\mathcal{O}(10$--$100$) nm are desired, the turbulent scales have to be much smaller, requiring an energy-intensive high-pressure homogenizer (HPH) method \citep{schultz2004high, haakansson2019emulsion}. 

Despite the wide range of scales involved, many key concepts crucial to understanding deformation and breakup in turbulence can be traced back to the seminal works by \citet{kolmogorov1949breakage} and \citet{hinze1955fundamentals}, i.e. the Kolmogorov-Hinze (KH) framework. The KH framework has gained widespread acceptance in various fields, however, it is crucial to acknowledge that it contains a number of assumptions and hypotheses. The purpose of this review is to gather studies from various disciplines that investigate the deformation and breakup of both droplets and bubbles in turbulence, in order to determine the regimes in which the KH framework is applicable and more importantly, where it may fall short and new challenges and opportunities await.

\begin{textbox}[h]
\label{box:ass}
\section{Key hypotheses and assumptions in the Kolmogorov-Hinze framework}
(a) Turbulence was assumed to be homogeneous and isotropic. (b) The drop size was assumed to be in the inertial range of turbulence. (c) Droplets were assumed to be neutrally buoyant, with buoyancy and density ratio disregarded. (d) It was hypothesized that the breakup is driven by the dynamic pressure caused by changes in velocity over distances at the most equal to the drop diameter. (e) The framework assumes that the interaction between drops and turbulence is one-way, with droplets having no effect on the turbulent dynamics. (f) While Kolmogorov took into account the 
kinematic viscosity ratio between the two phases to separate different regimes, Hinze proposed to use the Ohnesorge number (defined in Section 2) to measure the importance of the inner viscosity. 

 \end{textbox}

 This review provides an overview of the dynamics of deformation and breakup and their impacts on momentum, mass, and heat transfer, with a particular focus on experimental methods and results and a limited survey of simulation findings. For in-depth coverage of numerical methods for resolving deformable interfaces in turbulence, readers are referred to the recent reviews by \citet{tryggvason2013multiscale} and \citet{elghobashi2019direct}. The subject is closely related to the broader realm of particle-laden turbulence, including spherical \citep{balachandar2010turbulent, brandt2022particle}, non-spherical \citep{voth2017anisotropic}, and buoyant particles \citep{mathai2020bubble}, but with a particular emphasis on deformability. This review also complements other comprehensive reviews on fragmentation \citep{villermaux2007fragmentation}, bubble dynamics \citep{magnaudet2000motion,risso2018agitation,lohse2018bubble}, and the complexity introduced by surfactant \citep{takagi2011surfactant}, phase inversion \citep{bakhuis2021catastrophic}, and non-Newtonian liquids \citep{zenit2018hydrodynamic}.

The problem being considered involves bubbles and droplets of a specific diameter, denoted as $\sized$, being deformed and fragmented by surrounding turbulence characterized by parameters, such as energy dissipation rate ($\epsilon$), fluctuation velocity ($u'$), integral scale ($L$), and the Kolmogorov scale ($\eta$). The density, dynamic and kinematic viscosities are denoted by $\rho$, $\mu$, and $\nu$, respectively. The fluid properties of the carrier phase and dispersed phase can be differentiated using subscripts $c$ and $d$, respectively. The interfacial tension between the two phases is represented by $\sigma$.

The problem at hand is characterized by a multitude of parameters, and as a result, the relevant dimensionless groups are also vast. However, by making some key assumptions and hypotheses, as outlined in {\bf{Textbox \ref{box:ass}}}, \citet{kolmogorov1949breakage} was able to simplify the problem. He proposed that, for the deformation and breakup of large bubbles/droplets ($\eta\ll \sized\ll L$), the most important dimensionless number is the Weber number, which is a measure of the ratio between the inertial forces to surface tension forces.
\begin{equation}
    We_t=\frac{\rho_cu_D^2\sized}{\sigma}
    \label{eq:We}
\end{equation}
where $u_D$ is the eddy velocity of size $D$ and $u_D^2=C_2(\epsilon \sized)^{2/3}$ is the estimation using the second-order structure function in the inertial range in homogeneous and isotropic turbulence (HIT), where $C_2\approx 2.3$ is the Kolmogorov constant. Furthermore, it was postulated that, if the Weber number is the only dimensionless number that affects the breakup problem, there must exist a critical Weber number ($We_t^c$) that corresponds to the critical diameter($\sized^c$), below which the droplets remain stable for a prolonged period in turbulence.
\begin{equation}
    D^c=\left(\frac{We_t^c\sigma}{\rho_cC_2\epsilon^{2/3}}\right)^{3/5}
    \label{eq:sizeWe}
\end{equation}

The idea of a critical Weber number implies an abrupt shift from a finite breakup probability to zero at $We_t^c$, a simplistic view which does not account for turbulent fluctuations. Although the likelihood of eddies with local energy dissipation rates significantly higher than the mean is low, it is not zero. Therefore, even if the mean Weber number is below $We_t^c$, the occasional high-energy eddies can still break bubbles or droplets. Additionally, while $We_t$ captures the contribution of turbulence, persistent large-scale forcing, such as shear or buoyancy, can aid and even dominate deformation and breakup. To understand the fundamental breakup mechanisms, their contributions must be distinguished from those of turbulence. Lastly, incorporating the effects of viscosity poses significant challenges, requiring a systematic review of existing experimental data. To this end, this review is structured as follows. In Section 2, different regimes of deformation and breakup driven by turbulence, including the effects of large-scale forcing and viscous damping, are reviewed. Section 3 provides an overview of the key results and models of breakup frequency. In Section 4, the findings on how deformation and breakup influence the momentum, heat, and mass transfer between phases are summarized.

\section{Various breakup Regimes}

\begin{center}
\begin{figure}
    \includegraphics[width=\textwidth]{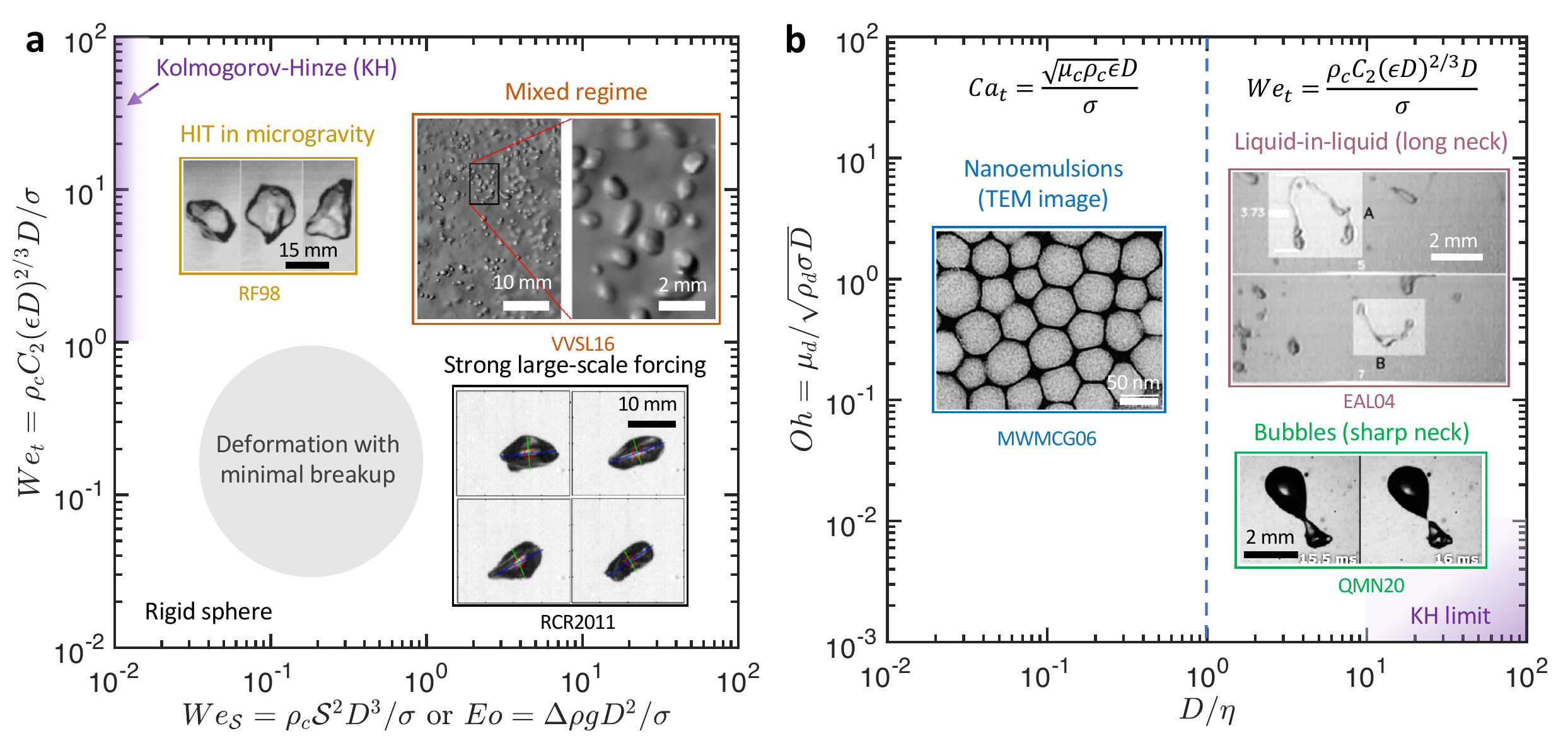}
    \caption{A parameter space of deformation and breakup of bubbles/droplets in turbulence characterized by (a) the Weber number defined based on the small-scale turbulence ($We_t$) versus large-scale persistent forcing measured by either shear ($We_{\mathcal{S}}$) or buoyancy ($Eo$), and (b) the Ohnesorge number ($Oh$) and the size of the bubbles/drops ($D$) relative to the Kolmogorov scale ($\eta$). The inset panels are adapted with permission from RF98 \citep{risso1998oscillations}, VVSL16 \citep{verschoof2016bubble}, RCR2011 \citep{ravelet2011dynamics}, MWMCG06 \citep{mason2006nanoemulsions}, EAL04 \citep{eastwood2004breakup}, and QMN20 \citep{qi2020towards}. The placement of these insets in the parameter space only indicates the general regimes they correspond to, not their exact parameters. }
    \label{fig:regime}
\end{figure}
\end{center}

{\bf{Figure \ref{fig:regime}}} illustrates the relevant regimes that have been studied and the typical deformation and breakup morphology that has been observed. {\bf{Figure \ref{fig:regime}}a} emphasizes the problems dominated by inertia but separately considers the effects of small-scale turbulence ($We_t$) and large-scale forcing. The large-scale forcing can arise in various forms, including a persistent mean shear with the shear rate denoted as $\mathcal{S}$ and a pressure gradient induced by buoyancy-driven migration, with their roles in deformation measured by the shear Weber number $We_{\mathcal{S}}=\rho_c \mathcal{S}^2 \sized^3/\sigma$ and the E\"{o}tv\"{o}s or Bond number $Eo=\Delta\rho g\sized^2/\sigma$, respectively.

{\bf{Figure \ref{fig:regime}}b} emphasizes the transition from an inertia-dominated to a viscous-dominated regime when $D$ crosses the Kolmogorov length scale ($\eta=(\nu_c^3/\epsilon)^{1/4}$) and the viscous effect becomes more pronounced. In the viscous regime, the crucial dimensionless number is the Capillary number, i.e. $Ca_t=\sqrt{\mu_c\rho_c\epsilon}\sized/\sigma$. As the viscous effect of the outer fluid becomes relevant, it is also necessary to consider the regimes when the inner viscosity matters as well. As a result, another key dimensionless number, i.e. the Ohnesorge number ($Oh=\mu_d/\sqrt{\rho_d\sigma D}$), is considered to measure the relative significance of $\mu_d$ in resisting and damping deformation.

\subsection{Inertia-dominated regime ($D>\eta$)}
\subsubsection{Intense homogeneous and isotropic turbulence ($We_t> Eo$ and $We_t> We_\mathcal{S}$)}

In this regime, the classical KH framework is most applicable. However, it can be difficult to achieve these ideal conditions in experiments. In closed systems, HIT can be generated by forcing flows from multiple symmetrical locations. As these flows merge, HIT can be produced near the center, where it is farthest from the momentum sources and therefore has the lowest energy dissipation rate. This location is also where measurements were typically taken. As a result, the probability of breakup is much higher outside the measurement volume than inside. Bubbles and drops that are likely to break would have already been broken before entering the measurement volume, making it challenging to study their behavior in classical HIT systems.



\begin{figure}[h!]
    \includegraphics[width=\textwidth]{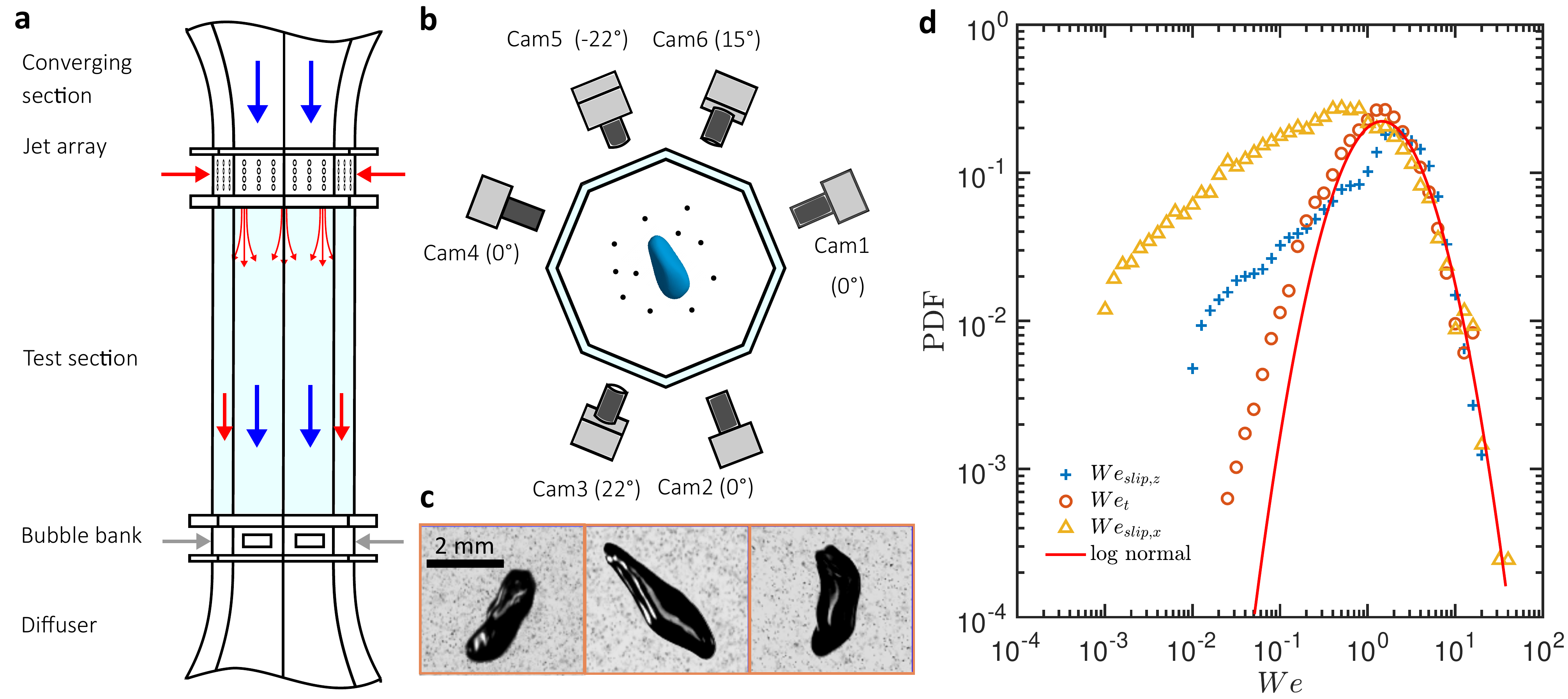}
    \caption{(a) Schematic of the side view of a vertical water tunnel that uses a jet array to produce intense HIT to study deformation and breakup \citep{masuk2019v,qi2020towards}. (b) The top view of the octagonal test section along with six cameras that were used to  measure the shape of deformed bubbles simultaneously with the nearby 3D turbulence \citep{masuk2019robust,qi2020towards}. (c) Examples of one strongly-deformed bubble captured by three different cameras. (d) The distribution of the Weber number defined based on different velocities. The red line indicates the log-normal distribution predicted based on the distribution of the instantaneous energy dissipation rate \citep{masuk2021simultaneous}. }
    \label{fig:setup}
\end{figure}

One solution is to use HIT that decays along one direction and guide bubbles or droplets through turbulence along the opposite direction. In this way, the energy dissipation rate that bubbles/drops encounter continues to increase and the measurement volume can be set at a location where the energy dissipation rate is the highest but the flow is still HIT.  \citet{masuk2019v} designed a vertical water tunnel with a jet array located at the top of the test section and firing jets co-axially with the mean flow downward into the test section. The facility and its key dimensions are shown in \textbf{Figure \ref{fig:setup}a}. \citet{Tan23jet} showed that, in this facility, the flow becomes HIT at around six nozzle spacings below the jet array, and such HIT continues to decay. The decay was found to scale with the nozzle diameter ($d_n$) and  the jet velocity at the nozzle exit ($v_j$). In particular, the fluctuation velocity follows $u'/v_j=(x/d_n)^{-1}$, and the energy dissipation rate decays as $\epsilon/(v_j^3/d)=0.76(x/d)^{-7/2}$. In this setup, the bubbles were injected at the bottom of the test section where the energy dissipation rate is the weakest. As they rise, the turbulence intensity grows, and eventually reaches a point where it is sufficient to cause bubbles to deform and break. Turbulence at this location, where the measurement volume is also placed, features large energy dissipation rates of $\mathcal{O}(1)$ m$^2$/s$^3$, which is sufficient for bubbles of size $\mathcal{O}(1)$ mm to reach the condition of $We_t>Eo$.

Apart from increasing turbulence intensity, another approach to reach $We_t>Eo$ is to weaken the buoyancy effect. \citet{risso1998oscillations} conducted experiments in a parabolic flight, reducing the gravitational constant to $g=0.4$ m/s$^2$ and effectively reducing $Eo$ by 25 times. In this experiment, turbulence was generated via an axisymmetrical momentum jet close to the bottom of the device \citep{risso1997diffusive}. The average size of bubbles investigated is about 18 mm, which would have broken due to buoyancy alone if the experiments were conducted under the Earth's gravity \citep{tripathi2015dynamics}, but under microgravity, the buoyancy effect was much weaker, and breakup was dominated by turbulence. One sequence of images for strongly-deformed bubble in such an environment is shown in an inset of {\bf{figure \ref{fig:regime}a}}. The critical Weber number averaged over all the cases was reported to be around 4.5. A numerical version of the same experiment was conducted by \citet{qian2006simulation} using the lattice Boltzmann method with bubbles being fragmented in homogeneous turbulence in a three-dimensional periodic box. The reported critical Weber number was around 3. 

\begin{marginnote}
\entry{Slip velocity}{the instantaneous velocity difference between the two phases. This quantity is different from the mean slip velocity used in atomization.}
\end{marginnote}

Although microgravity helps reduce the impact of the relative motion between the two phases (slip velocity) driven by buoyancy, this rising motion is not the sole source of the additional pressure gradient that could drive deformation. Even with a neutrally-buoyant dispersed phase, as assumed in the KH framework, slip velocity ($u_{\text{slip}}$) cannot be fully eliminated due to finite size effects \citep{homann2010finite,bellani2012slip}. To determine the extent to which deformation is driven by slip velocity, \citet{masuk2021simultaneous} conducted an experiment to measure the shape of deforming bubbles simultaneously with their surrounding turbulence in 3D. This challenging experiment was accomplished using a diagnostic setup, which included six cameras positioned around the test section ({\bf{figure \ref{fig:setup}b}}). Typical images of a deformed bubble, along with the nearby tracers, are shown in {\bf{figure \ref{fig:setup}c}}. The shadows of high concentration of tracers were tracked using the openLPT method \citep{tan2020introducing}, while the bubble geometry was reconstructed using a technique that employs surface tension as an additional physical constraint, resulting in improved reconstruction quality \citep{masuk2019robust}.


These simultaneous measurements provide insight into the Lagrangian evolution of the bubble Weber number and the shape of individual bubbles \citep{masuk2021simultaneous}. The distributions of the Weber number based on different velocity scales are shown in {\bf{figure \ref{fig:setup}d}}. $We_{t}=\rho(\widetilde{\lambda_3}D)^2D/\sigma$, in this case, was determined by using the eigenvalue, $\lambda_3$, that corresponds to the most compressive direction ($\hat{\mathbf{e}}_3$), of the strain rate tensor coarse-grained at the bubble scale. The ensemble average of this definition of $We_t$ should be equivalent to the one proposed in the KH framework, but it provides a more accurate representation of the relevant instantaneous Weber number, which was confirmed by the fact that the semi-minor axis of the bubble preferentially aligns with $\hat{\mathbf{e}}_3$. This alignment suggests that it is the converging flow near the bubble and the resulting pressure rise on the interface that leads to compression and deformation. The PDF of $We_t$ can be captured through a log-normal distribution (red line in {\bf{figure \ref{fig:setup}d}}), calculated by following the definition of $We_t$ and the distribution of local energy dissipation rate that is described by the refined Kolmogorov theory \citep{kolmogorov1962refinement} and the multi-fractal spectrum \citep{meneveau1991multifractal}.

In addition to the strain rate, the instantaneous slip velocity was also calculated and divided into the horizontal ($x$) and vertical components ($z$). Their respective Weber numbers, i.e. $We_{\text{slip,x}}$ and $We_{\text{slip,z}}$, can be calculated by using the slip velocity as the velocity scale. The PDF of $We_{\text{slip,z}}$ contains the contribution by the buoyancy-driven deformation, but the overall shapes of $We_{\text{slip,x}}$ and $We_{\text{slip,z}}$ remain similar to each other. The difference between $We_{\text{slip,x}}$ and $We_t$ ({\bf{figure \ref{fig:setup}d}}), on the other hand, is significant, underscoring the difference between two deformation mechanisms brought by turbulence and finite-sized bubbles.

Through simultaneous measurements, a simple relationship between $We_t$ and aspect ratio of the bubble $\alpha$, following $\alpha = 2We_t^{2/3}/5 + 1.2$, was determined by minimizing the difference between the PDF of $\alpha$ obtained from direct shape measurement and that calculated from $We_t$. This relationship provides a way to describe bubble deformation in turbulence, which complements studies that investigated the deformation of gas bubbles rising in quiescent liquids \citep{legendre2012deformation}. In turbulence, while it was noted that the fit against $We_t$ is slightly better than against $We_{\text{slip}}$, the difference was not substantial, suggesting that both are important in bubble deformation. However, the orientation analysis by \citet{masuk2021orientational} indicates that the bubble semi-minor axis aligns signficantly better with the slip velocity than with $\hat{\mathbf{e}}_3$, emphasizing the crucial roles played by slip velocity in bubble deformation in turbulence.

\subsubsection{Shear or buoyancy dominated deformation ($We_t< Eo$ or $We_t< We_\mathcal{S}$)}
\paragraph{Buoyancy dominated regime ($We_t< Eo$)}



In this regime, deformation is predominantly driven by buoyancy and the turbulence effect is minimal. \citet{sevik1973splitting} conducted an experiment on the breakup of bubbles in a turbulent jet. Bubbles with diameter varying from 4.0 mm to 5.8 mm were injected along the centerline of the jet. $Eo$ ranges from 2.1 to 4.5, and the critical Weber number determined was about 1.3. This critical $We_t$ is much smaller than 4.5 reported by \citet{risso1998oscillations} based on the experiments conducted in microgravity, indicating less stress from turbulence was needed to break bubbles thanks to the extra help from buoyancy.

\citet{ravelet2011dynamics} conducted an experiment of large bubbles rising in weak turbulence and reported two different Weber numbers, one based on the bubble's typical rise velocity and the other on the velocity gradient across the bubble. The fact that the former Weber number was close to 11.6 indicates that bubbles were strongly deformed due to buoyancy. The latter one, $We_t$ as defined in equation \ref{eq:We}, was around 1.8, which was about an order of magnitude smaller. {\bf{Figure \ref{fig:regime}a}} displays snapshots of a deforming bubble, which show resemblance to bubbles rising in a still medium \citep{mougin2001path}, with the short axis of the bubble preferentially tilted towards the vertical direction and exhibiting periodic motions. These similarities were expected since bubbles were still primarily compressed in the vertical direction and the same wake instability occurred \citep{zenit2008path}. Despite these similarities, the time series of bubble deformation in turbulence were more chaotic and the decorrelation timescale was associated with the mode-2 natural frequency of the small-amplitude bubble oscillation, i.e. $f_{2}=\sqrt{96 \sigma/\rho_c D^{3}}$. The natural oscillation timescale was proposed as an important timescale by \citet{sevik1973splitting} and \citet{risso1998oscillations} based on the physical picture of a bubble resonating with turbulent perturbations at its natural frequency. However, with strong buoyancy, \citet{ravelet2011dynamics} suggested that the preferential sliding motion between the two phases significantly changes the deformation dynamics, leading to breakup driven by single intense eddies rather than the stochastic resonance observed under microgravity ($We_t>Eo$). This work implied that persistent deformation in one direction could alter the deformation dynamics driven by turbulence more than just adding to the stress.


\paragraph{Shear dominated regime ($We_t< We_\mathcal{S}$)}
\citet{levich1962physicochemical} considered the breakup of small drops immersed in the logarithmic sub-layer of a turbulent boundary layer (TBL). The mean velocity parallel to the wall, $\langle U\rangle$, in the wall normal direction ($y$) is described by $\langle U\rangle=U_{\tau}\ln(y/y_0)/\kappa$, where $U_{\tau}=\sqrt{\tau_w/\rho_c}$ is the friction velocity and is determined by the wall shear stress ($\tau_w$). The characteristic length scale is expressed as $\delta_0=\nu_c/U_{\tau}$ or $y_0=\delta_0/9$. $\kappa\approx 0.4$ is the von-K\'arm\'an constant. \citet{levich1962physicochemical} argued that the pressure gradient that drives the drop deformation is dominated by the persistent large-scale shear across the drop size $D$ from $y$ to $y+D$. Assuming $d\ll y$, the Weber number ($We_\mathcal{S}$) can be expressed as a function of $y$
\begin{equation}
    We_\mathcal{S}=\frac{2\rho_c U_{\tau}^2\sized^2}{\kappa^2y\sigma}\ln\frac{y}{y_0}
    \quad\mathrm{and}\quad 
        We_\mathcal{S,\text{max}}=\frac{\ln(180)\rho_c U_{\tau}^3\sized^2}{10\kappa^2\nu_c\sigma}\approx \frac{3\rho_c U_{\tau}^3\sized^2}{\kappa^2\nu_c\sigma}
    \label{eq:we_levich1}
\end{equation}
where $We_\mathcal{S,\text{max}}$ is the largest value of $We_\mathcal{S}$ that can be reached near the bottom end of the log layer ($y\approx 20\delta_0$). Equation \ref{eq:we_levich1} is slightly different from the original work by \citet{levich1962physicochemical} after I corrected some issues, e.g. the assumption of the bottom of the log layer at $y\approx e\delta_0$. Although it was not explicitly mentioned in the work by \citet{levich1962physicochemical}, $We_\mathcal{S,\text{max}}$ can be re-written as $We_\mathcal{S,\text{max}}=3Re_{d}\tau_w \sized/\kappa^2\sigma$, which is essentially the Weber number based on the wall shear stress ($We_\tau=\tau_w \sized/\sigma$) multiplied by the droplet Reynolds number $Re_c=U_{\tau}\sized/\nu_c$ based on $U_{\tau}$ and the carrier-phase fluid properties. 

\citet{yi2021global} studied the behavior of an oil-water emulsion in a Taylor-Couette (TC) system, which consists of a fluid layer between two counter-rotating cylinders. The resulting flows featured two thin turbulent boundary layers (TBLs) near the surfaces of the inner and outer cylinders, leaving a larger bulk region with more homogeneous and isotropic turbulence. \citet{yi2021global} first employed the KH framework, assuming that most of the droplet breakup occurred within the bulk region. Although the scaling law was verified, as the results were reanalyzed \citep{yi2022physical,yi2023recent}, it became evident that the value of the critical Weber number is much smaller than 1, from 0.013 to 0.018 using the bulk turbulence, implying that the turbulent stresses were not sufficient to overcome the interfacial tension in the bulk. This led them to conclude that the majority of droplets stayed in the bulk but most breakup occurred in the TBLs. When using Levich's definition of Weber number, \citet{yi2022physical} found that the critical Weber number is close to 5, which is order unity and more reasonable. This finding suggests that the breakup was indeed driven primarily by the mean shear in the TBL, whose thickness was around 5 times the average droplet diameter.

Bubble breakup also occurs when they are directly injected into the near-wall region of a TBL. \citet{madavan1985measurements} found that the bubble size in the TBL is determined by the free-stream velocity and gas flow rate, and is not affected by the method of gas injection. This finding implies that the size distribution of bubbles is primarily controlled by breakup and coalescence. Rather than following equation \ref{eq:we_levich1}, \citet{pal1988bubble} proposed a new way to calculate the Weber number by estimating the local energy dissipation rate $\epsilon\sim U_\tau^3/\theta$ ($\theta$ is the momentum thickness) experienced by bubbles based on the turbulence within the boundary layer. \citet{sanders2006bubble} revised this definition, replacing the momentum thickness $\theta$ with $\kappa y$, resulting in a new definition of the turbulent Weber number.
\begin{equation}
    We_t=\frac{2\rho_cU_\tau^2D^{5/3}}{(\kappa y)^{2/3}\sigma}
    \label{eq:WeSt}
\end{equation}
Bubbles located approximately $y=$1 mm ($y\approx370\delta_0$) away from the wall in flows with a friction velocity of $U_\tau=0.37$ m/s have been observed to have a size distribution of  320$\pm$130 $\mu$m \citep{sanders2006bubble}. This size distribution can be explained by assuming $We_t$ in equation \ref{eq:WeSt} has a critical value. But it is important to note that, for a critical $We_t$ near unity, $We_\mathcal{S}$ using equation \ref{eq:we_levich1} is roughly 19.

\subsubsection{Mixed regime ($We_t\approx We_\mathcal{S}$)}

Most turbulent flows, e.g. homogeneous shear flow \citep{rosti2019droplets,Ferrante2023}, turbulent pipe or channel flows \citep{angeli2000drop, scarbolo2015coalescence,mangani2022influence}, breaking waves \citep{garrett2000connection,deane2002scale},  turbulent jets \citep{martinez1999breakup1}, von K\'{a}rm\'{a}n swirling flow \citep{ravichandar2022turbulent}, or stirred tanks 
 \citep{shinnar1961behaviour}, typically involve both large-scale flows and turbulence.

 Efforts have been made to minimize the impact of large-scale flows by injecting bubbles or droplets in regions that are closer to HIT, such as the centerline of jets \citep{martinez1999breakup1} or the center of the von K\'{a}rm\'{a}n swirling flow \citep{ravichandar2022turbulent}. However, as the injected bubbles or droplets are carried away from the injection point, the influence of large-scale flows may still be present.

In chemical or petroleum engineering, the size of oil droplets broken by turbulence in pipes or stirred tanks is often studied using the critical Weber number defined based on the global energy dissipation rate $\langle\epsilon\rangle$, where $\langle...\rangle$ represents the ensemble average over the entire device. For batch rotor-stator systems and conventional stirred tanks, $\langle\epsilon\rangle$ scales with the rotor speed $N$ and the  rotor diameter $L$, as expressed by $\langle\epsilon\rangle\sim N^3 L^2$ \citep{rushton1950power, chen1967drop}. This $\langle\epsilon\rangle$ leads to the critical Weber number being defined as $We^c_t=\rho_cN^2L^{4/3}\sized^{5/3}/\sigma$, from which the critical drop size can then be determined.


In pipe flows, the global energy dissipation rate is expressed as $\langle\epsilon\rangle=fU_c^3/2D_p$, where $f$ is the friction factor, $U_c$ is the mean axial velocity of the continuous phase, and $D_p$ is the pipe diameter. This leads to a critical Weber number that scales with $Df^{2/3}$. However, \citet{kubie1977drop} argued that the velocity scale should be the fluctuation velocity, not the centerline velocity. The fluctuation velocity is approximately equal to 1.3$U_{\tau}$, where $U_{\tau}=(f/8)^{1/2}U_c$. Based on this, the critical Weber number can be rewritten as $We^c_t=f\rho_c\sized U_c^2/\sigma$, which scales with $f$ instead of $f^{2/3}$. However, experiments conducted by \citet{angeli2000drop} found that the critical drop size scales with $f^{-3}$, albeit from a very narrow range of $f$. Thus, more experiments are needed to fully resolve the debate and determine the appropriate relationship between the critical drop size and friction factor in pipe flows.


\subsubsection{Viscous effects in the inertia-dominated regime}
\label{sec:vis}
For droplets with size $D\gg\eta$, the deformation is dominated by flow inertia and the viscosity of the carrier phase can be neglected. However, the inner viscous damping may still play a significant role in deformation dynamics, comparable or even exceeding the impact of surface tension, as quantified by the dimensionless number, $Oh$.

 \citet{davies1985drop} and \citet{calabrese1986drop} considered this problem and assumed a total balance of stresses between the external forcing by turbulence and internal damping: $\rho_c\left(\varepsilon \sized\right)^{2 / 3}\sim\sigma/ \sized+\mu_{\mathrm{d}}\left(\varepsilon \sized\right)^{1 / 3} \sqrt{\rho_c/\rho_d} / \sized$. It can also be expressed in the dimensionless form, $We_t\sim c_1+c_2Oh^2Re_d\sqrt{\rho_c/\rho_d}$, where $Re_d=\rho_d(\epsilon D)^{1/3}D/\mu_d$ is the droplet Reynolds number based on the eddy velocity and inner fluid properties, and $c_1$ and $c_2$ are two fitting constants. Droplets will deform if the left side is  larger than the right side, which implies that quantities of interest $\mathcal{Q}$ (e.g. aspect ratio or breakup frequency) should be a function of the new dimensionless number following 
\begin{equation}
\mathcal{Q}=f\left( \frac{We_t}{c_1+c_2Oh^2Re_d\sqrt{\rho_c/\rho_d}}\right)
    \label{eq:weoh}
\end{equation}
Equation \ref{eq:weoh} indicates that, for deformation and breakup, the primary dimensionless number is $We_t$ when the inner viscous damping is negligible, and $We_t/Oh^2Re_d\sqrt{\rho_c/\rho_d}$ when it is important to consider.




To investigate the viscous effect, \citet{eastwood2004breakup} injected oil droplets in a turbulent jet along the centerline, using the same setup as \citet{martinez1999breakup1} in their investigation of bubble breakup. The values of $Oh$ in the experiment ranged from $\mathcal{O}(10^{-2})$ to $\mathcal{O}(10^{-1})$, and $We_t$ was roughly $\mathcal{O}(10)$. As illustrated in {\bf{figure \ref{fig:regime}}b}, the experiment revealed a clear long filament, indicating a significant deformation preceding breakup. The extent of stretching increased with an increase in droplet viscosity, and droplets within the inertial sub-range stretched to lengths comparable to the local integral scale before fragmentation. This long filament was confirmed by other experiments \citep{andersson2006breakup, solsvik2015single} and simulations \citep{haakansson2022criterion}, and it was found to be connected to the large number of daughter droplets generated from droplet breakup. 

Recognizing the importance of this process, \citet{maass2012determination} adopted the idea originally proposed by \citet{janssen1993droplet} to describe a drop elongating in one dimension and thinning in the other two exponentially over time, driven by a constant straining flow. Given a stretching rate, the breakup time could be estimated once a critical diameter of the neck was determined, which was assumed to be related to the critical Capillary number based on $\mu_d$ and the stretching rate. However, this model did not account for the instability of the filament itself \citep{ruth2022experimental} or the possible interruption by small-scale eddies, as it assumed a persistent elongation at the scale of the filament.

\citet{vankova2007emulsification} investigated the size of emulsion droplets produced using a HPH with various oils, resulting in a range of $Oh$ from $\mathcal{O}(10^{-1})$ to $\mathcal{O}(10)$. The authors adopted equation \ref{eq:weoh} and adjusted two constants, $c_1$ and $c_2$, to fit their experimental results. The obtained values were $0.78$ and $0.37$ for $c_1$ and $c_2$ respectively. However, subsequent analysis by \citet{Zhong2023} questioned the validity of the linear combination of the restoring and dissipative terms in equation \ref{eq:weoh} and proposed a new equation to better collapse all the data.

\begin{equation}
    \mathcal{Q}=f\left(\frac{We_t}{1+Oh}\right)
    \label{eq:weohni}
\end{equation}
$\mathcal{Q}$, in this case, is the non-dimensionalized breakup frequency. This quantity will be further discussed in detail in Section \ref{sec:breakfreq}. Note that this relationship was established based on limited experimental data. In order to further examine the validity of this relationship, it is possible to use simulation databases, such as the one by \citet{mangani2022influence}, with well-controlled characteristics that cover a broad range of density and viscosity ratios in turbulence.



\subsection{Viscous-dominated regime ($\sized<\eta$)}


\subsubsection{Experimental Methods}

There are three main experimental methods for producing droplets with $\sized<\eta$: (a) generating turbulence with high $\epsilon$; the typical value of $\epsilon$ can be calculated by satisfying two criteria $Ca_t=\sqrt{\rho_c\mu_c\epsilon}D/\sigma>Ca_t^c$ and $D<(\nu_c^3/\epsilon)^{1/4}$. (b) reducing surface tension by adding surfactants, and (c) increasing the viscosity of the carrier phase. In food processing industries, including dairy, breaking droplets into nanometer sizes is important for the desired texture, color, and stability for storage. Option (b) or (c) is less ideal due to the required food-grade chemical additives, leaving option (a) as the primary method. The HPH is the key technique for this purpose \citep{haakansson2019emulsion}. It uses a high-pressure piston pump (50--200 MPa) and a narrow gap ($\mathcal{O}(100)$ $ \mu$m) to accelerate emulsions to velocities of up to $\mathcal{O}(100)$ m/s, creating a localized turbulent jet \citep{bisten2016optical} with $\epsilon$ in the range of $\mathcal{O}(10^8)$ to $\mathcal{O}(10^9)$ m$^2$/s$^3$ and fragmenting droplets to sizes of $\mathcal{O}(10$--$100)$ nm.

The high-speed colloid mill (rotor-stator) system, is another commonly used method for emulsion preparation. This type of system is similar to the high-Reynolds-number TC system \citep{van2011twente, grossmann2016high}, but with a small gap size ($h=\mathcal{O}(100$) $\mu$m) and high rotor spin rate (inner cylinder with the radius of $r_i=\mathcal{O}(10$) cm) at $\omega_i=\mathcal{O}(10^4$) revolutions per minute (RPM), which results in a moderate Reynolds number ($Re=\omega_i r_i h / \nu_c$) at around $\mathcal{O}(10^4$) but a significant mean shear and turbulent energy dissipation rate \citep{schuster2012analysis}.

\subsubsection{Negligible inner viscosity $Oh\ll 1$}

By systematically increasing the viscosity of the continuous oil phase ($\mu_c$) by two orders of magnitude while keeping the dispersed aqueous phase constant (option c in the previous section), \citet{boxall2012droplet} studied the transition of the dynamics of droplet breakup from the inertia-dominated to the viscous-dominated regimes. The droplets were fragmented by turbulence in a customized mixing cell driven by a six-blade impeller. The droplet size was determined using the focused beam reflectance method, and the average droplet size was calculated only after the steady state was reached, which took approximately three hours.


If $Oh$ is negligibly small, the only dimensionless number that matters to the problem is the capillary number ($Ca_t$). Assuming that a critical capillary number ($Ca_t^c$) exists, \citet{shinnar1961behaviour} suggested that the critical droplet size ($D^c$) can be determined as follows
 \begin{equation}
    D^c=\frac{Ca_t^c\sigma}{\sqrt{\mu_c\rho_c\epsilon}}
     \label{eq:sizeCa}
 \end{equation}
In the experiments conducted by \citet{boxall2012droplet}, the impeller speed ($N$) and diameter ($L$) were kept almost constant, so the energy dissipation rate ($\epsilon\sim N^3L^2$) did not vary significantly. As $\mu_c$ increased, it was shown that the droplet size remained unchanged for low values of $\mu_c$, and scaled with $\mu_c^{-1/2}$ in the viscous-dominated regime at high $\mu_c$. This finding provides direct support for Equation \ref{eq:sizeCa}.



\begin{figure}

    \includegraphics[width=1\textwidth]{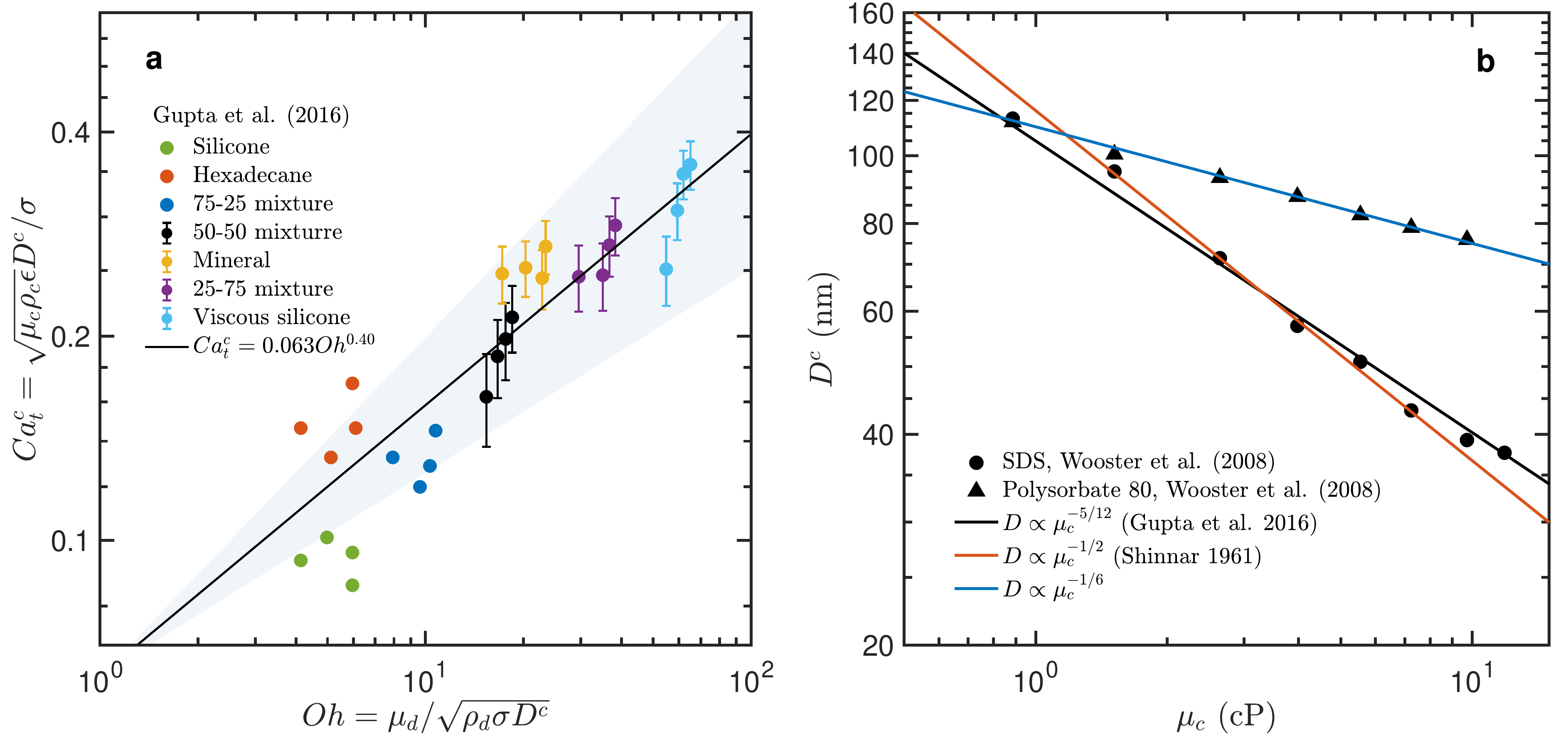}
    \caption{(a) The critical Capillary number measured from nanoemulsions made with a homogenizer as a function of $Oh$. The black solid line shows the 2/5 scaling, with the blue shaded area representing the $\pm 0.1$ uncertainty in the scaling exponent. (b) The critical drop size $D^c$ versus the viscosity of the carrier phase for nanoemulsions made with two different types of surfactants. Figures are adapted with permission from \citet{wooster2008impact} and \citet{gupta2016controlling}}
    \label{fig:gupta}
\end{figure}

\subsubsection{Large inner viscosity $Oh\gtrsim1$}
In the previous section, a water-in-oil emulsion was considered and the viscosity of the dispersed phase was negligible in comparison to the continuous phase. For other types of emulsions, such as oil-in-water, $\mu_d$ is large and the viscous damping by the inner fluid cannot be neglected, and it is likely that $Ca_t^c$, if it exists, is dependent on the value of $Oh$.

To model this dependence, \citet{gupta2016nanoemulsions} proposed a model based on the physical picture of a part of the droplet, with the size of the instability length scale, being extruded from the parent droplet due to the surrounding turbulence. By assuming that the propagation timescale of the instability is dominated by the viscous diffusion of eddy momentum into the droplet, a new formulation was derived, $Ca^c_t\sim Oh^{2/5}$, indicating that the critical Capillary number is not a constant, but a function of $Oh^{2/5}$. {\bf{Figure \ref{fig:gupta}a}} shows the measured $Ca^c_t$ over a range of $Oh$ by systematically varying the types of oils used in experiments. The oil droplet size obtained from HPH was substituted into the definition of $Ca_t$ to obtain its value. The proposed scaling seems to capture the scaling between $Ca^c_t$ and $Oh$ well.

This model proposed by \citet{gupta2016nanoemulsions} further predicts the critical droplet size relative to all other fluid properties as follows: $\sized^c=C_1(\sigma^{5/6}\mu_d^{1/3})/[(\rho_d\sigma)^{1/5}(\mu_c\rho_c\epsilon)^{5/12}]$. Specifically, it implies that $\sized$ scales with $\mu_d^{1/3}$, $\mu_c^{-5/12}$, and $\epsilon^{-5/12}$. The scaling was compared to experimental data obtained by \citet{wooster2008impact}, who created an oil-in-water emulsion with varying $\mu_c$ by adding various concentrations of polyethylene glycol into water. The comparison is shown in {\bf{figure \ref{fig:gupta}b}}. Although the proposed scaling of $D\sim\mu_c^{-5/12}$ by \citet{gupta2016controlling} agrees well with the data, it is difficult to distinguish it from the one proposed by \citet{shinnar1961behaviour} ($D\sim\mu_c^{-1/2}$), which did not account for $\mu_d$ and $Oh$. 

In addition, \citet{wooster2008impact} actually reported two datasets using the same emulsions with the only difference being the types of surfactant added. The data adopted by \citet{gupta2016controlling} is the one with 5.6 wt \% sodium dodecylsulphate (SDS, 98.5\%). The data using 5.6 wt \% polysorbate 80 (Tween 80, 98\%) is also shown in {\bf{figure \ref{fig:gupta}b}} (triangles), which deviates noticeably from the proposed -5/12 scaling, agreeing better with $D\sim\mu_c^{-1/6}$. This  difference implies the possible complexity introduced by surfactant and coalescence.


Nevertheless, assuming the scaling proposed by \citet{gupta2016controlling} is correct and combining the two regimes, i.e. with or without significant viscous effects, one can express quantities of interest in the problem of viscous deformation and breakup using an equation similar to equation \ref{eq:weoh}
\begin{equation}
\mathcal{Q}=f\left( \frac{Ca_t}{c_3+c_4Oh^{2/5 }}\right)
    \label{eq:caoh}
\end{equation}
where $c_3$ and $c_4$ are constants that are yet to be determined to understand the critical Capillary number and the transitional $Oh$ from the regime where the inner viscous damping is important to where it is not.

\section{Deformation and breakup: timescales and dynamics}

\subsection{Characteristic timescales}

The prediction of the evolution of the drop and bubble size distribution over space and time can be captured by solving the population balance equation, which is a Boltzmann-type equation. This approach has been widely implemented in many simulation methods to predict the dynamics of polydispersed particles, bubbles, and drops that constantly coalesce or break \citep{marchisio2013computational,shiea2020numerical}. In the population balance equation, there are three key quantities to describe breakup, the breakup frequency, the daughter bubble/droplet size distribution, and the number of daughters. 


For breakup frequency, selecting the right timescale to non-dimensionalize it is the first challenge. The discussions of the characteristic timescale of deformation and breakup can be traced back to Section 127 of the book by \citet{levich1962physicochemical}, who considered four different breakup timescales based on the inner viscosity and interface velocity. These regimes can be determined by estimating the magnitude of the three terms, the pressure gradient $\nabla p/\rho_d$, unsteady term $\partial u_d/\partial t$, and viscous term $\nu_d\nabla^2 u_d$, in the wave equation that describes the inner fluid motion during breakup. Four timescales have been proposed, as expressed in the following equations. 
\begin{equation}
    \frac{\partial u_d}{\partial t}\sim\frac{\nabla p}{\rho_d}\sim \frac{p}{D\rho_d};\frac{\partial u_d}{\partial t}\sim\frac{D}{\tau^2}; p\sim\frac{\sigma}{D}\Rightarrow\tau\sim\sqrt{\frac{\rho_dD^3}{\sigma}}~(\textrm{low viscosity, low speed})
     \label{eqn:timeA}
\end{equation}
\begin{equation}
    \frac{\nabla p}{\rho_d}\sim\nu_d\nabla^2 u_d\sim \frac{\nu_d}{D\tau};\frac{\nabla p}{\rho_d}\sim \frac{p}{D\rho_d};p\sim\frac{\sigma}{D}\Rightarrow\tau\sim\frac{\mu_dD}{\sigma}~(\textrm{high viscosity, low speed})
        \label{eqn:timeB}
\end{equation}
\begin{equation}
    \frac{\partial u_d}{\partial t}\sim\frac{\nabla p}{\rho_d}\sim \frac{p}{D\rho_d};\frac{\partial u_d}{\partial t}\sim\frac{D}{\tau^2}; p\sim\rho_c u_c^2\Rightarrow\tau\sim\frac{D}{u_c}\sqrt{\frac{\rho_d}{\rho_c}}~~(\textrm{low viscosity, high speed})
        \label{eqn:timeC}
\end{equation}
\begin{equation}
    \frac{\nabla p}{\rho_d}\sim\nu_d\nabla^2 u_d\sim \frac{\nu_d}{D\tau};\frac{\nabla p}{\rho_d}\sim \frac{p}{D\rho_d}; p\sim\rho_c u_c^2\Rightarrow\tau\sim\frac{\mu_d}{\rho_c u_c^2}~~(\textrm{high viscosity, high speed})
        \label{eqn:timeD}
\end{equation}
where $\tau$ is the characteristic breakup timescale. $u_d$ is the characteristic inner fluid velocity, which does not show up in the final estimation of $\tau$ because $u_d$ scales roughly with $D/\tau$.

The eddy turnover time has been proposed as another characteristic timescale, $t_D=\epsilon^{1/3} D^{-2/3}$, for describing bubble fragmentation in breaking waves \citep{garrett2000connection,deane2002scale,chan2021turbulent,gao2021bubble}. In particular, the scaling between $t_D$ and $D$ directly results in the steady-state bubble size distribution scaling with $D^{-10/3}$, which was also observed in droplet breakup in turbulence \citep{soligo2019breakage,crialesi2023interaction}. Note that the eddy turnover time is, in fact, in line with equation \ref{eqn:timeC} given by \citet{levich1962physicochemical}, if the characteristic velocity scale ($u_c$) of the outer flow is set as the eddy velocity at the bubble size $u_D=(\epsilon D)^{1/3}$, as proposed in the KH framework. The only difference left is that $t_D$ does not account for the density ratio between the two phases.

Another proposed timescale is the natural oscillation frequency, $f_2$, which is associated with the second eigenmode of weak-amplitude oscillations \citep{lamb1879hydrodynamics}. Assuming inviscid fluids, $1/f_2=\sqrt{(3\rho_d+2\rho_c)D^3/30\sigma}$. Although similar to the timescale listed in equation \ref{eqn:timeA}, there is an important distinction to note: Levich's model only accounted for the density of the inner fluid, whereas a more complicated relationship with both $\rho_d$ and $\rho_c$ is required for $1/f_2$.


\subsection{Experimental results}
\begin{figure}
    \includegraphics[width=\textwidth]{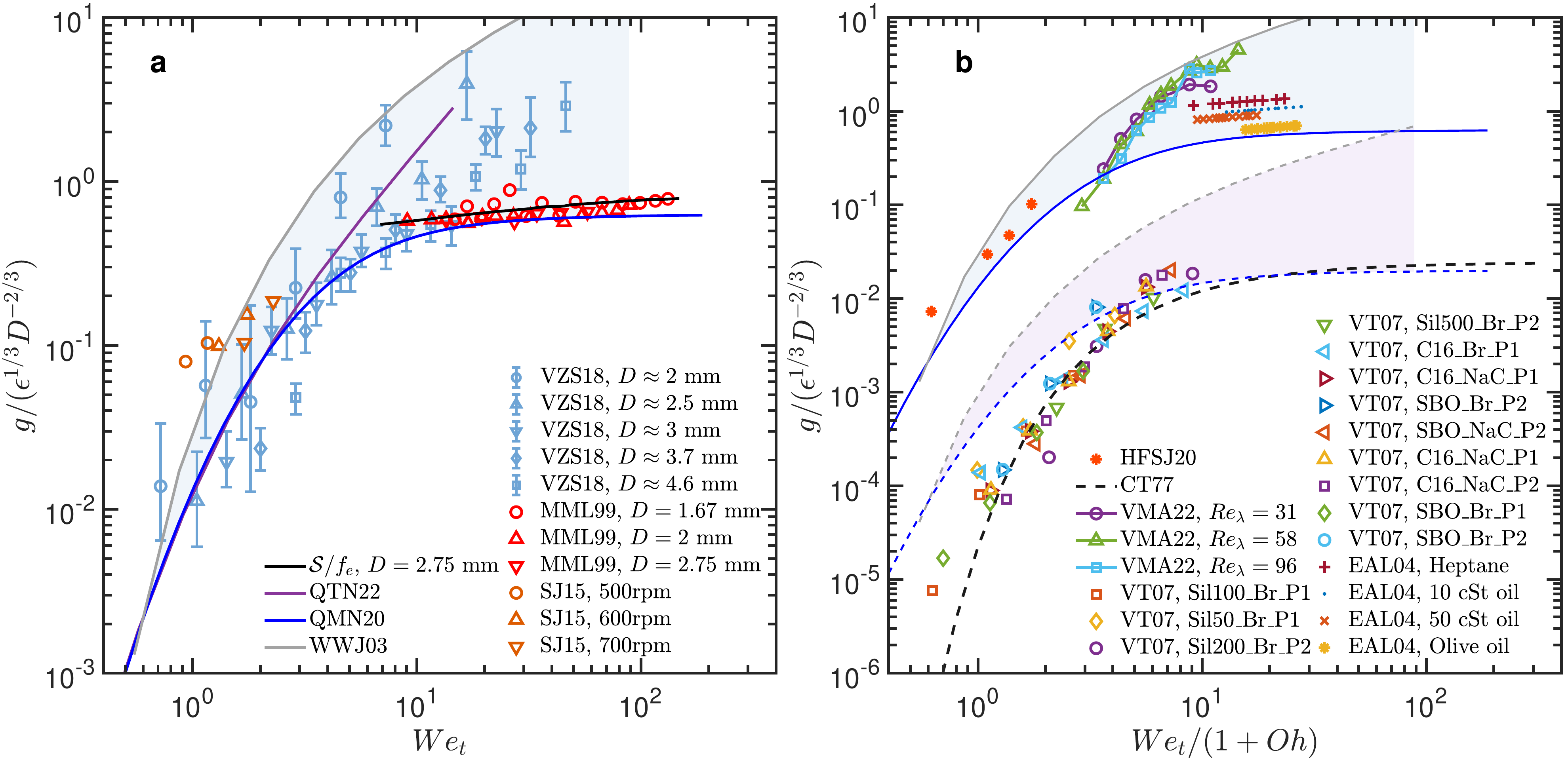}
    \caption{Breakup frequency of (a) bubbles and (b) droplets normalized by the eddy turnover frequency, $f_e=(\epsilon^{1/3}D^{-2/3})$, as a function of the key dimensionless number. The datasets that were compiled include VZS18 \citep{vejrazka2018}, MML99 \citep{martinez1999breakup1}, SJ15 \citep{solsvik2015single}, VT07\citep{vankova2007emulsification}, EAL04 \citep{eastwood2004breakup}, VMA22 \citep{vela2022memoryless}, HFSJ20 \citep{hero2020single}. Models include the ones by CT77 \citep{coulaloglou1977description}, QTN22 \citep{qi2022fragmentation},QMN20 \citep{qi2020towards}, and WWJ03 \citep{wang2003novel}. 
    }
    \label{fig:breakfreq}
\end{figure}

\label{sec:breakfreq}

\citet{Zhong2023} compiled experimental results on the breakup frequency of bubbles and oil droplets with sizes $D>\eta$ based on the recommendation made by \citet{haakansson2020validity}. The eddy turnover time of size $\sized$ was used as the characteristic timescale, chosen from a list of options mentioned in the previous section. {\bf{Figure \ref{fig:breakfreq}}} clearly shows that the breakup frequency drops sharply as $We_t$ decreases, indicating droplets or bubbles with smaller $We_t$ take longer to break. Thus, it is evident that the definition of a critical Weber number depends on the observation time \citep{vela2022memoryless}. If one waits longer during an experiment, smaller droplets or bubbles can be obtained for a given level of turbulence.

Although the data for bubbles ({\bf{figure \ref{fig:breakfreq}}a}) showed better agreement, discrepancies were still noticeable at high $We_t$. Specifically, \citet{martinez1999breakup1} reported a plateau in breakup frequency ($g$) close to 1, while \citet{vejrazka2018} claimed that $g$ increased towards 10 without reaching a plateau. This disparity could be due to the different experimental conditions employed: the former was conducted in turbulence closer to HIT, while the latter involved injecting bubbles along the centerline of a turbulent jet. As the bubbles spread away from the centerline as they migrate downstream, they may experience mean shear.

\begin{marginnote}
\entry{Breakup frequency ($g$)}{the fraction of bubbles/droplets  that break per unit time in turbulence}
\end{marginnote}

The inverse of the mean shear rate ($\mathcal{S}$) was suggested by \citet{Zhong2023} as another potential timescale, estimated by dividing the centerline velocity of the turbulent jet by its width, following the canonical turbulent jet. The estimated shear rate is plotted as a black solid line in {\bf{figure \ref{fig:breakfreq}}a}, matching the measured breakup frequency quite well and providing an alternative timescale to consider for breakup frequency in inhomogeneous and anisotropic turbulence.

For droplet data in {\bf{figure \ref{fig:breakfreq}}b}, equation \ref{eq:weohni} mentioned in Section \ref{sec:vis}, proposed by \citet{Zhong2023}, was used to compile different datasets with different $Oh$ . The datasets provided by \citet{vankova2007emulsification}, with $Oh$ ranging from $\mathcal{O}(10^{-2})$ to $\mathcal{O}(10)$, collapsed with one another well using equation \ref{eq:weohni}. However, when including all the datasets, a nearly two-orders-of-magnitude variation in the breakup frequency was observed. This difference primarily arises from the experiments conducted by \citet{vankova2007emulsification} in a homogenizer, where the drop size was in the order of $\mathcal{O}(10^{-5})$ m, as compared to other experiments that involved much larger drops of $\mathcal{O}(10^{-3})$ m, implying either large systematic uncertainties between large- and small-scale experiments or a potential hidden size dependence that was not accounted for in the current selection of dimensionless groups.

The deformation and breakup of bubbles and droplets could potentially be understood under a unified framework if a suitable set of dimensionless numbers is chosen to collapse all available data. In an attempt to achieve this, \citet{Zhong2023} selected two models that represent the upper \citep{wang2003novel} and lower \citep{qi2020towards} bounds of bubble experiments, as illustrated by the shaded area in Figure \ref{fig:breakfreq}a. The same shaded area was overlaid twice on top of the droplet datasets in Figure \ref{fig:breakfreq}b, once with (lower shaded area with dashed lines as bounds) and once without (upper shaded area with solid lines as bounds) the density ratio, $\sqrt{\rho_d/\rho_c}$, as suggested by Levich's timescales (Equation \ref{eqn:timeC}), to show how well the bubble and droplet data collapse. The inclusion of $\sqrt{\rho_d/\rho_c}$ resulted in the collapse of most of the available bubble and droplet data, except for the dataset by \citet{vankova2007emulsification}. In contrast, when the density ratio was not considered, the bubble data showed better agreement with the results by \citet{vankova2007emulsification}. This finding suggests that the existing droplet data exhibits too much disparity to draw a definitive conclusion regarding the effectiveness of including the density ratio term for characterizing the breakup timescale.

\subsection{Numerical simulations} 

In addition to experiments, with the development of more advanced direct numerical simulation (DNS) algorithms for two-phase flows \citep{elghobashi2019direct} and Graphics Processing Unit (GPU) based codes \citep{crialesi2023flutas}, it is possible to conduct a large number of simulations of breakup events to collect statistics. For example, \citet{liu2021efficient} implemented an efficient simulation scheme for the phase-field method to simulate the breakup of a large drop and the coalescence of $\mathcal{O}(10^3)$ drops. 

In addition to the simulation schemes, in general, two strategies have been adopted so far. The first one involves a larger simulation domain with many drops and a limited number of selected dimensionless numbers, and drops are allowed to break and coalesce \citep{dodd2016interaction,roccon2017viscosity,scarbolo2015coalescence,mangani2022influence,crialesi2023interaction}. This strategy is particularly suitable for investigating breakup at high concentrations in complex environments that are relevant to many applications, such as emulsions and breaking waves. Since it simulates both breakup and coalescence in turbulence, it also helps illustrate the energy transferred between the two phases \citep{dodd2016interaction,crialesi2023interaction}. 

The second method relies on a smaller domain with only one drop but many more runs, ranging from hundreds \citep{riviere2021sub} to over 30,000 \citep{vela2022memoryless}, to cover a wider parameter space. The advantage of this method is the isolation of the breakup events without the complication of coalescence. This approach is particularly useful for investigating parameters under which breakup takes a long time, a regime where experiments suffer from large uncertainty and finite residence time.

\subsection{Deformation and breakup models} 
If reliable models for describing deformation and breakup can be developed, it is much more computationally efficient to integrate these models along the Lagrangian trajectories of point bubbles/droplets in turbulence to study their breakup frequency and probability. In the following, we review some of the deformation and breakup models.

 For viscous fluids, \citet{maffettone1998equation} developed a model (M\&M) to describe the evolution of both shape and orientation of neutrally-buoyant spheroidal droplets in a linear velocity gradient. This model was validated against several experimental studies in low Reynolds number. Recently, the model has been applied to simulating the deformation of many sub-Kolmogorov-scale neutrally-buoyant droplets ($D\ll\eta$) in turbulence by integrating the M\&M equation numerically along their Lagrangian trajectories \citep{biferale2014deformation, spandan2016deformation}.

For inertia-dominated deformation and breakup, it is much more challenging to develop a deformation model. One model simplified the problem by ignoring the orientation and proposed to describe a droplet as a linear damped oscillator that is forced by the instantaneous turbulent fluctuations at the drop scale \citep{risso1998oscillations,lalanne2019model}. The equation can be written in a dimensionless form as follows,
\begin{equation}
\label{eq:risso}
\frac{d^2\hat{a}}{d\hat{t}^2}+2\xi\frac{d\hat{a}}{d\hat{t}}+\hat{a}=K'We_t(t)
\end{equation}
where $\hat{a}$ represents the difference between the semi-major axis of the deformed geometry and the spherical-equivalent radius divided by $D/2$. The damping coefficient is given by $\xi=1/2\pi\tau_d f_2$, where $\tau_d$ is the damping time scale defined as $\tau_d = D^2 / 80 \nu_c$ for bubbles \citep{risso1998oscillations} but a much more complicated implicit solution for droplets \citep{lalanne2019model}. In contrast to previous models that assumed an additive relationship between viscous stress and surface tension \citep{davies1985drop, calabrese1986drop}, this model correctly incorporated the dissipative nature of viscous damping, and it has been successfully compared to experimental data on breakup statistics, even in inhomogeneous turbulence \citep{galinat2007dynamics,maniero2012modeling}.



To account for the orientation and add multiple deformation mechanisms, \citet{masuk2021towards} adapted the MnM model into the inertia-dominated regime by making three important modifications: (a) Velocity gradients were coarse-grained at the size of the bubble; (b) Deformation due to slip velocity was accounted for by using a pseudo-strain-rate tensor; and (c) A pseudo-rotation tensor was added to model the wake-induced bubble rotation. The modified model has been successfully used to predict deformation and orientation for bubbles in both turbulence and quiescent media, with the predicted statistics agreeing well with experimental data. These findings suggest that the modified model effectively captures the key mechanisms responsible for inertial deformation and breakup.

\subsection{Recent models for breakup frequency}
Recent advances in modeling breakup mechanisms have highlighted the importance of several previously neglected factors, including gas density, eddies of different sizes, and turbulence intermittency, which we will summarize here. In the past, air bubbles were often modeled as having negligible density and viscosity. However, it has been shown that bubbles made of heavier gases can break more frequently \citep{wilkinson1993influence}. This phenomenon was explained by \citet{andersson2006breakup}, who pointed out that deformation typically results in a dumbbell shape with two uneven ends. As the smaller end retracts due to surface tension, air flow accelerates through the neck, which reduces the local pressure and speeds up the breakup process. Larger gas density tends to lower the local pressure and shorten the breakup time even further. These observations and proposed mechanisms have inspired new models developed by \citet{xing2015unified} and \citet{zhang2020improved}, which incorporate the effect of backflow and gas density.

In addition to the density effect, to accurately model bubble-eddy collision, it is crucial to determine which eddy scales should be considered. The KH framework assumes that only the drop-scale eddy is significant, with both larger and smaller scales being negligible. Conversely, some models consider eddies of all length scales from the turbulent spectrum \citep{karimi2018exploratory, castellano2019using}. However, recent research by \citet{vela2021deformation}, which employed direct numerical simulation of a single drop being deformed in turbulence, found that the impact of eddies with different length scales on the variation of surface free energy is not equal. Turbulent fluctuations at scales smaller than the drop diameter cause the majority of surface deformation, while the contribution of scales close to or larger than $D$ is relatively minor.


\citet{qi2022fragmentation} designed an experiment using the head-on collision between two vortex rings to isolate the turbulent scales. During the early stage before the collision, only intact large-scale vortices were accessible, while the post-collision late stage was filled with many small eddies. Despite a lower overall $We_t$ in the late stage, bubbles were found to break up in a more violent and faster manner due to the presence of small eddies. Building on this finding, the authors developed a new model that considers not only the stress criterion, which requires the incoming eddy to exert sufficient stress to overcome the restoring surface tension, but also the time scale. The breakup must occur within the time before the bubble relaxes. This key idea emphasizes that, instead of being gradually and consistently stretched by flows at their own length scales, bubbles are fragmented by small eddies, resulting in a sudden and intense local deformation over a short period of time. The predicted breakup frequency as a function of $We_t$ is shown as the purple solid line in {\bf{figure \ref{fig:breakfreq}a}}, which agrees with the experimental data by \citet{vejrazka2018}.



Numerous studies have investigated the effect of turbulence intermittency on bubble breakup. Recent models have examined the impact of intermittency on the turbulent energy spectrum, as noted by \citet{bagkeris2021modeling} and \citet{solsvik2016review}. However, the effect of intermittency on the distribution of $\epsilon$, which can be derived from the multi-fractal model and described by a log-normal distribution \citep{meneveau1991multifractal}, is more pronounced than that on the energy spectrum. This distribution can be incorporated into modeling quantities such as the breakup probability \citep{masuk2021simultaneous}, eddy velocity \citep{qi2022fragmentation}, and breakup frequency \citep{qi2020towards}. 

In particular, \citet{qi2020towards} modified the model originally proposed by \citet{martinez1999breakup1} to account for the non-negligible breakup frequencies for small bubbles when exposed to intermittent turbulent eddies. The model prediction is shown as the blue solid line in {\bf{figure \ref{fig:breakfreq}a}} and the lower bounds for the two shaded areas in {\bf{figure \ref{fig:breakfreq}b}}. The classical model by \citet{coulaloglou1977description} (black dashed line) fits the data by \citet{vankova2007emulsification} well. However, for most other datasets, a slower decay of the breakup frequency $g$ as $We_t$ decreases is observed, which is better predicted by \citet{qi2020towards}.

\section{Modulation of mass, momentum, and heat transfer by deformation}

\subsection{Deformation affecting effective bubble forces}
The motion of large bubbles and droplets in turbulence can be characterized by the combined effect of multiple hydrodynamic forces, such as buoyancy, drag, lift, added mass, Basset history, and pressure forces \citep{magnaudet2000motion, sridhar1995drag}. As the majority of these forces are shape-dependent, it is not surprising that bubble and droplet deformability can significantly impact their translational motion and local concentration in turbulence.

Most research on forces experienced by bubbles has focused on their behavior in laminar shear \citep{legendre1998lift, tomiyama2002transverse, lu2008effect, dijkhuizen2010numerical, hessenkemper2020contamination}. Bubble deformation, driven primarily by buoyancy, is measured by $Eo$. As $Eo$ increases, the lift force undergoes a transition from positive to negative values. This shift in direction is attributed to the stretching and tilting of vorticity generated at the bubble surface, which transforms into a pair of counter-rotating streamwise vortices in the bubble wake. These vortices have the opposite sign compared to those produced around a spherical bubble, resulting in a negative lift force. In addition to vorticity production, direct asymmetric deformation caused by external shear can also lead to negative lift \citep{zhang2021three, hidman2022lift}. 


In turbulence, \citet{sugrue2017robust} proposed a new dimensionless number taking the product of $Eo$ and the ratio between the local turbulent kinetic energy and the squared relative velocity between the two phases. This new number is linked to the Weber number based on the fluctuation velocity. The authors carried out extensive simulations, varying lift coefficients, and minimizing the differences between experimental and simulated results. This allowed them to extract lift coefficients for different flow conditions. The results showed that the lift coefficients exhibited a similar inversion to those observed in laminar shear flow. However, two key differences were noted. Firstly, the magnitude of the coefficients was much smaller, and secondly, the inversion diameter was smaller for turbulence-driven cases.

To measure the lift coefficient in turbulence, \citet{salibindla2020lift} conducted an experiment in nearly HIT. Although the flow does not have a mean shear, the bubbles were constantly subjected to local shear and vorticity. The transition of bubble rising velocity in turbulence from lower to faster than its counterpart in an otherwise-quiescent medium was found as the bubble size increased. Based on this finding and the access to the statistics of both phases, the lift and drag forces experienced by bubbles with different sizes were determined, and the lift inversion at smaller bubble size was observed experimentally. The lift inversion was correlated to the turbulence-induced deformation measured by $We_t$, which is close to 1 as the inversion occurs, suggesting that turbulence-induced bubble deformation becomes essential. The transition of the bubble's rising velocity was linked to the preferential sampling of different regions (upward or downward) in turbulence. This work also supports the mechanism proposed by \citet{spelt1997motion} that small spherical bubbles tend to preferentially sample the downward flows in turbulence instead of being trapped in the vortex cores \citep{wang1993motion} and also quantitatively explains other previous experiments \citep{poorte2002experiments,aliseda2011preferential,prakash2012gravity}.

It is worth noting that the transition of bubble rising velocity was not observed in another work conducted by \citet{ruth2021effect}. In this study, the change in rise velocity was attributed mainly to drag rather than lift, highlighting the need for further investigation into how deformable bubbles modify lift and drag forces in intense turbulence where deformation is driven by local turbulence instead of buoyancy. Nevertheless, once lift and drag forces are determined, the added mass force can also be evaluated experimentally. Recent work by \citet{salibindla2021experimental} showed that the added mass force experienced by bubbles in turbulence can be accurately modeled using the solid spheroid approximations \citep{lamb1879hydrodynamics}. These findings suggest that the instantaneous added-mass force experienced by deformable bubbles can be approximated by appropriately oriented spheroids with the correct instantaneous aspect ratios.



\subsection{Turbulent drag reduction}
\begin{figure}
    \includegraphics[width=\textwidth]{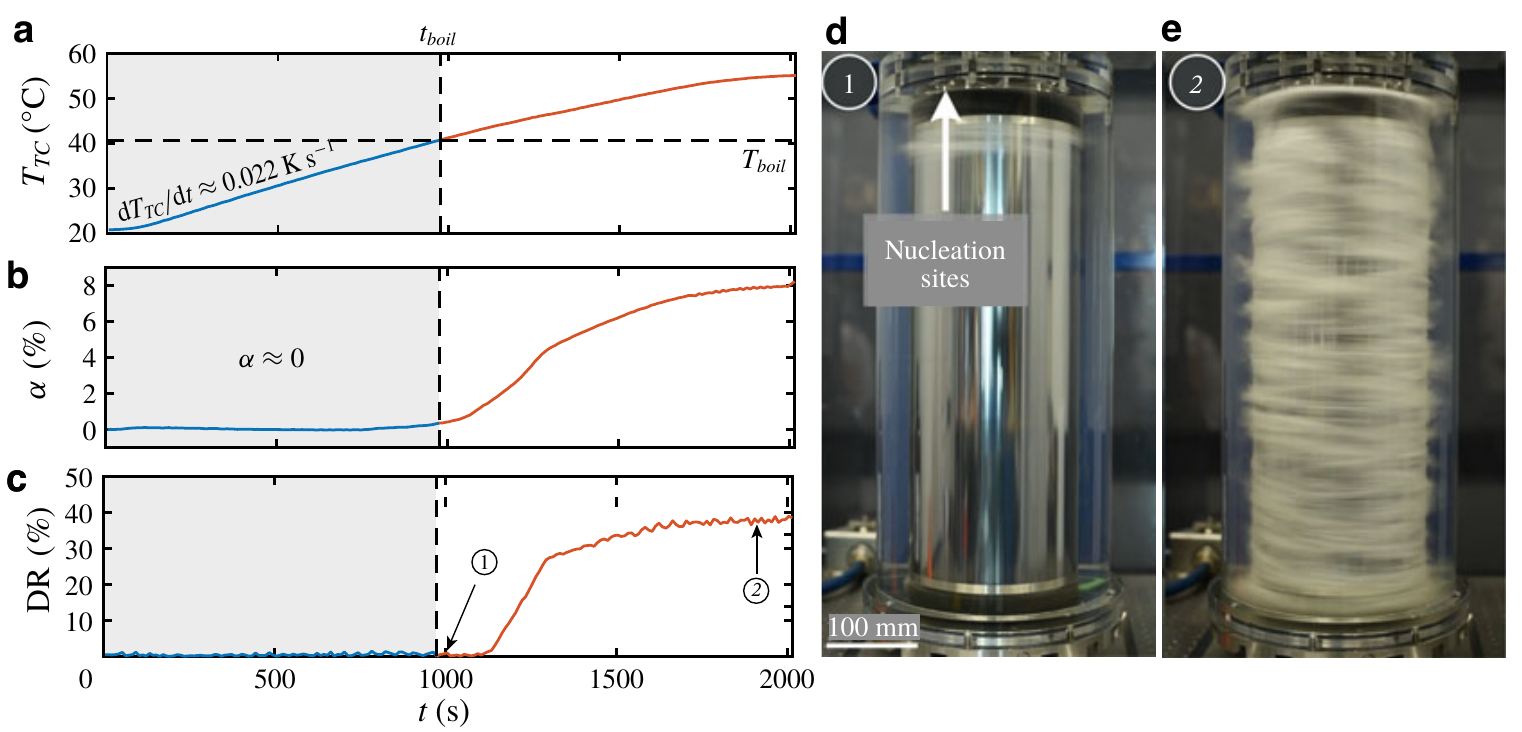}
    \caption{The drag of turbulent Taylor–Couette flow during its transition from non-boiling (grey shaded areas) to boiling at $t=t_{boil}$ and the key quantities, including (a) Liquid temperature $T_{TC}$, (b) volume
fraction $\alpha$, (c) drag reduction (DR) as a function of time; Two time steps correspond to the photographs
shown in (d) and (e). Figures are adapted with permission from \citet{ezeta2019drag}.}
    \label{fig:drag}
\end{figure}

The deformation and breakup of bubbles and drops in turbulent boundary layers have been extensively studied in the context of drag reduction \citep{ceccio2010friction, murai2014frictional} in various configurations, including turbulent TC \citep{van2013importance}, flat plates \citep{sanders2006bubble}, channel flows \citep{murai2007skin, tanaka2021repetitive}, and even under the model ship hull \citep{tanaka2022frictional}. Several mechanisms have been proposed to explain the origin of the bubble-induced drag reduction effect \citep{ferrante2004physical,lu2005effect,lu2008effect,van2007bubbly}. The successful drag reduction experiments have been summarized by \citet{murai2014frictional} in two regimes: relatively small bubbles in high-speed flows or large bubbles in low-speed flows. A recent overview of this topic was presented by \citet{lohse2018bubble}, who highlights the difference between bubble-induced drag reduction at small Reynolds numbers and large Reynolds numbers, attributed to the effects of Froude number and Weber number, respectively.

While most experiments on bubble-mediated drag reduction focused on large-scale averaged skin friction, a few studies measured the couplings between the two phases. \citet{kitagawa2005flow} conducted 2D simultaneous measurements of both phases in a horizontal turbulent channel to determine the mechanism of drag reduction caused by bubbles. The bubbles, which were roughly 530 $\mu$m and deformable close to the wall, were shown to cause a drop in the Reynolds stress of the carrier phase. The reduction ratio was almost the same as that of the skin friction coefficient. \citet{murai2007skin} measured the fluctuation of the wall shear stress due to the passage of individual deformable bubbles with size comparable to the boundary layer thickness using a shear transducer. They found that the bubbles considerably reduced the local wall shear stress, which was induced by the two roll vortices upstream and downstream of the bubble that modified the local turbulent shear stress \citep{oishi2014horizontal}.

Drag reduction has been a topic of extensive study in the context of high-Reynolds-number TC systems because it is a closed system where frictional drag can be measured as the global torque. \citet{van2013importance} showed that the system transitions from moderate drag reduction of 7\% to a more significant one with nearly 40\% drag reduction at Reynolds number ($Re$) above $10^6$ and gas void fraction of 4\%. This transition was observed as the Weber number crosses one and the bubble becomes more deformable, even as the size becomes smaller as $Re$ increases. As $Re$ increases and drag decreases, a larger bubble aspect ratio was observed (as shown in {\bf{figure \ref{fig:regime}a}}), signaling the connection between deformation and drag reduction. This point was further supported by \citet{verschoof2016bubble}, who showed that the large drag reduction (40\%) could be `turned off' by adding some surfactant. The surfactant reduces surface tension and hinders coalescence, which leads to much smaller bubbles with smaller $We_t$. Similar levels of drag reduction were also observed in boiling TC driven by vapor bubbles, again due to their deformation in turbulence, as shown in {\bf{figure \ref{fig:drag}}}. As the vapor bubble volume fraction $\alpha$ increases, the probability of finding a larger value of $We_t$ increases, probably due to the presence of more-deformable larger bubbles formed by coalescence. Finally, a recent study by \citet{wang2022turbulence} investigated how viscosity ratios between the two phases affect drag in TC and found that the drag coefficient increases as the inner viscosity increases and drop deformability weakens, further reaffirming the importance of deformation in turbulent drag reduction.


Extensive simulations have been also conducted to explore the potential mechanism of drag reduction driven by deformable bubbles/droplets. \citet{iwasaki2001direct} demonstrated that the droplets can attenuate near-wall streamwise vortices via deformation. \citet{lu2005effect} found that large deformable bubbles can lead to significant drag reduction by suppressing streamwise vorticity near the wall, whereas less-deformed bubbles tend to bring additional shear rate near the viscous sublayer to increase drag. \citet{spandan2018physical} reported that deformable bubbles can reduce drag in TC flows by modulating dissipation in their wakes, regardless of whether the carrier fluid is weakly or highly turbulent. Overall, these studies underscore different mechanisms at play in bubble/droplet-mediated turbulent drag reduction through a deformable interface.

\subsection{Turbulence modulation}

\citet{dodd2016interaction} performed direct numerical simulations to investigate the behavior of finite-sized drops in decaying isotropic turbulence, exploring a range of Weber numbers, density ratios, and viscosity ratios between the two phases. In this work, the turbulence kinetic energy (TKE) equations were derived to capture the energy transfer between two phases, with a particular focus on the role of interfacial energy. It was shown that, while the presence of droplets always enhances the dissipation rate near the droplet interface, the initial turbulence decay rate is faster in the presence of more deformable drops (i.e., larger Weber numbers). However, the decay rate becomes independent of the Weber number later on, likely because the turbulence has decayed to a point where it is no longer strong enough to deform or break any drops. The study also demonstrated that droplet coalescence acts as a source of TKE through the power of surface tension, while breakup serves as a sink of TKE.

In their investigation of turbulence modulation at different scales, \citet{freund2019wavelet}  analyzed the same data generated by 
 \citet{dodd2016interaction} using wavelet transforms instead of Fourier transforms because wavelet restricts the effects of non-smoothness locally while preserving spatial information. At a distance larger than 5$\eta$ or $D/4$, the carrier-phase spectra remained nearly unaffected, but the energy at high wavenumbers increased close to the drop interface due to enhanced local velocity gradients. They also observed that drops with larger density ratios reduced the energy at low wavenumbers compared to neutrally-buoyant drops. In a separate study, \citet{scarbolo2013unified} examined the interaction between turbulence and one large deformable droplet and showed that the
presence of the interface results in vorticity generation and turbulence damping near the interface, and the distance from the interface where these effects are present depends on the surface tension.

The spectrum analysis of turbulence modulation by drops in HIT was also studied by 
\citet{mukherjee2019droplet}. They showed that the presence of dispersed drops leads to a transfer of energy from large scales to small scales, as the drops subtract energy from the former and inject it into the latter. This transfer of energy is reflected in the energy spectra, which cross the spectra of the single-phase turbulence at a length scale close to the Kolmogorov-Hinze scale, as initially proposed by \citet{perlekar2014spinodal} for a different system. \citet{crialesi2022modulation} provided further insights into the mechanisms behind this phenomenon, showing that surface tension forces play a key role in absorbing energy from large scales and reducing its transfer through advection terms. Eventually, this energy is transferred to small scales by surface tension. They also noted that the modulation of turbulence spectra is more sensitive to the viscosity ratio, while the scale-by-scale energy budget depends more on the volume fractions.  

Bubble-induced turbulence in a swarm of rising bubbles in an otherwise quiescent fluid has been extensively studied in recent years. The phenomenon has been investigated \citep{riboux2010experimental, innocenti2021direct,pandey2022turbulence}, modelled \citep{ma2017direct,du2019analysis}, and also recently reviewed by \citet{risso2018agitation} and \citet{mathai2020bubble}. \citet{mercado2010bubble} measured the energy spectrum of the carrier phase using a phase-sensitive anemometry and found that the energy decays with wave number following a power law of $-3.2$, which is close to the $-3$ scaling proposed by  \citet{lance1991turbulence}. They also observed that even at a small gas volume fraction, typically from 0.28\% to 0.74\%, deformable bubbles tend to cluster along the vertical direction at both small and large scales, which was attributed to deformation \citep{bunner2003effect}. 


\subsection{Heat and mass transfer}
The study of heat and mass transfer in turbulent multiphase flows is a complex and multifaceted topic that deserves a dedicated review because a wide range of relevant interfacial transfer phenomena, including boiling/condensation \citep{ russo2014water}, dissolution \citep{mac2015shape,farsoiya2023direct}, melting \citep{machicoane2013melting}, evaporation \citep{birouk2006current,duret2012dns,marie2014lagrangian,mees2020statistical}, and the induced Stefan flow, can be potentially modulated by turbulence and a deformable interface.

Deformation and breakup have been shown to affect heat and mass transfer. However, the extent of their influence on these processes is not yet fully understood beyond their effect on the size distribution and interfacial area. Recently, \citet{albernaz2017droplet} investigated the deformation and heat transfer of a single drop in HIT and found a negative correlation between local curvature and temperature on the droplet surface. \citet{wang2019self} found that the kinematics of deformable bubbles and droplets could significantly enhance the heat transfer in turbulent convection, and revealed that the emergent size distribution of the bubbles and droplets in the system governed the degree of heat transfer enhancement achievable. \citet{dodd2021analysis} used direct numerical simulation (DNS) to study finite-size, deformable, and evaporating droplets in HIT, and they showed that higher surface curvature induced by deformation and breakup leads to higher evaporation rates, especially for cases with large Weber numbers. \citet{shao2022interaction} demonstrated that the Stefan flow induced by evaporation reduces the coalescence rate and attenuates the turbulence kinetic energy. \citet{scapin2022finite} extended the problem to homogeneous shear flow and found that the larger surface area due to deformation leads to an overall larger mass transfer rate for drops with higher Weber numbers in persistent mean shear. They also observed a weak correlation between the interfacial mass flux and curvature at high temperature and a positive correlation at large Weber number, low ambient temperature, and slower evaporation. \citet{boyd2023consistent} simulated the aerodynamic breakup of an acetone drop in a high-speed and high-temperature vapor stream and showed that as the drop deforms, the increase of frontal surface area results in a significantly increased rate of evaporation and a nonlinear decrease in drop volume over time.

\begin{summary}[SUMMARY POINTS]
\begin{enumerate}

\item New experiments capable of measuring the shape of deforming bubbles and drops simultaneously with the detailed surrounding turbulence in 3D have been made possible with the advancement of diagnostic methods and new facilities that can generate controlled turbulence.
\item
Turbulence in many applications is usually inhomogeneous and anisotropic. Deformation and breakup in these systems are often subjected to both a non-uniform distribution of turbulence intensity and a persistent large-scale shear. 

\item 

The primary dimensionless parameter for breakup driven by the flow inertia of the carrier phase ($D\gg\eta$) is $We/(1+Oh)$, while for breakup driven by the viscous stress of the carrier phase ($D\ll\eta$), it is $Ca/(c_3+c_4Oh^{2/5})$. These relationships were established based on limited data, and further studies are required to validate them. 

\item
Deformation is driven by large-scale eddies and breakup is accelerated by small-scale eddies. The multiscale nature of breakup is a key component in understanding the inertia-dominated breakup. 

\item Existing works showed the intricacies of the interplay between the local interface curvature, local interfacial mass flux, and the induced Stefan flow for two-phase heat and mass transfer with a deformable interface. 

\end{enumerate}
\end{summary}

\begin{issues}[FUTURE ISSUES]
\begin{enumerate}


\item
Understanding the potential synergies between large-scale shear and local turbulence in deforming and breaking bubbles/droplets requires additional research.

\item
The impact of density ratio, Reynolds number, and additives such as salt and surfactant on deformation and breakup should be further investigated, as their roles remain poorly understood despite previous research.

\item
Additional research is needed to better understand how the time history of bubbles or droplets affects their deformation and breakup processes as they traverse through inhomogeneous turbulence with varying local intensities.

\end{enumerate}
\end{issues}

\section*{DISCLOSURE STATEMENT}
The author is not aware of any affiliations, memberships, funding, or financial holdings that might be perceived as affecting the objectivity of this review. 

\section*{ACKNOWLEDGMENTS}
The author thanks colleagues who provided advice and suggestions on the manuscript. The author also acknowledges support from NSF grant CBET 1854475 and CAREER-1905103 and Office of Naval Research grant N00014-21-1-2083 and N00014-21-1-2123.

\bibliographystyle{ar-style1.bst}
\bibliography{review_V6.bbl}

\end{document}